\documentstyle[aas2pp4,natbib209]{article}

\begin{document}

\title{Colliding Decimeter Dust}

\author{J. Deckers\altaffilmark{1} and J. Teiser}
\affil{Fakult\"at f\"ur Physik, Universit\"at Duisburg-Essen, 47057 Duisburg}

\altaffiltext{1}{johannes.deckers@uni-due.de}

\begin{abstract}
Collisional evolution is a key process in planetesimal formation and decimeter bodies play a key role in the different models. However, the outcome of collisions between two dusty decimeter bodies has never been studied experimentally. Therefore, we carried out microgravity collision experiments in the Bremen drop tower. The agglomerates consist of quartz with irregularly shaped micrometer-sized grains and the mean volume filling factor is 0.437 $\pm$ 0.004. The aggregates are cylindrical with 12 cm in height and 12 cm in diameter and typical masses are 1.5 kg. These are the largest and most massive dust aggregates studied in collisions to date. We observed rebound and fragmentation but no sticking in the velocity range  between 0.8 and 25.7 cm s$^{-1}$. The critical fragmentation velocity for split up of an aggregate is 16.2 $\pm$ 0.4 cm s$^{-1}$. At lower velocities the aggregates bounce off each other. In this velocity range, the coefficient of restitution decreases with increasing collision velocity from 0.8 to 0.3. While the aggregates are very weak, the critical specific kinetic energy for fragmentation $Q_{\mu =1}$ is a factor six larger than expected. Collisions of large bodies in protoplanetary disks are supposed to be much faster and the generation of smaller fragments is likely. In planetary rings collision velocities are of the order of a few cm s$^{-1}$ and are thereby in the same range investigated in these experiments. The coefficient of restitution of dust agglomerates and regolith covered ice particles, which are common in planetary rings, are similar.
\end{abstract}


\section{Introduction}
Planets form in protoplanetary disks by accretion of kilometer sized protoplanetary bodies, commonly called planetesimals. This later growth phase is driven by gravity  \citep{Wetherill2000, Wetherill1993}. The planetesimals themselves grow from dust particles, but due to the small size and mass, self gravity is not important during 
individual collisions and the basic formation process -- if a single one can be named -- is not yet understood entirely.\\
The different growth models can roughly be divided into two groups. One group considers growth of planetesimals through aggregation of dust particles by mutual collisions. Many studies showed that aggregation is very efficient for the growth from micron-sized particles to millimeter-sized aggregates \citep{blum2008, Dullemond2005, Krause2004, suyama2012, wada2009, seizinger2012}. Collision velocities in this size range are small (few mm s$^{-1}$). This leads to hit-and-stick collisions where surface forces provide the glue for sticking. Once particles reach the millimeter sized range, compaction of the particles leads to a more complex impact behavior. For compact dust agglomerates bouncing and fragmentation are common results. Bouncing might prevent further growth under certain conditions \citep{Zsom2010,guettler2010}.\\
Fragmentation and bouncing reduce the growth probability for mutual collisions between dust granules of similar size. Experiments with dust agglomerates of different sizes show that larger aggregates (centimeter to decimeter) can collect small (100 $\mu$m) particles as they drift through a reservoir of small grains at high velocity \citep{wurm2005, teiser2009, teiser2011}. In this kind of collision the small particles fragment and a great part of the kinetic energy is dissipated. Part of the original particle mass then sticks to the surface of the target body, while fragments are ejected at velocities of about 10\% of the original impact velocity. \citet{teiser2009} and \citet{wurm2001} showed that impact ejecta can be re-accreted by the target body, as they are accelerated by gas drag back to the surface.\\
It was shown that decimeter aggregates can grow by sweeping up small particles and that this process leads to a universal volume filling factor (ratio of dust to void volume within an aggregate) of the growing dust agglomerates of $\phi$ = 0.32 \citep{teiser2011}.\\
\\ \citet{windmark2012a} showed that with an existing bouncing barrier (at mm-size) and some larger seeds capable of sweeping up smaller particles the formation of planetesimals 
is possible. Larger seeds might be provided by several ways \citep{windmark2012b}. \citet{jankowski2012} find that mm-size aggregates consisting of larger monomers $\sim$ 10 $\mu$m still show a probability for sticking. \citet{weidling2012} find that once a couple of larger granules of dust are stuck together this aggregates of granules might grow in collisions at larger velocities.\\
Other possibilities of seeding include processes based on the radial drift of particles from the outer parts of protoplanetary disks into the inner parts \citep{weidenschilling1977}. If large icy aggregates form more easily in the outer regions, sublimating during drift through the snowline might provide larger dust aggregates as seeds as well \citep{saito2011, aumatell2011}. In total, growth of planetesimals through mutual collisions seems to be one viable mechanism for growing planetesimals.\\ 
In detail, the particle evolution depends on the outcome of individual collisions of all kinds of sizes. For small sizes, a growing number of experiments provide input \citep{guettler2010}. For larger aggregates, the database of collisional outcomes is rather poor. There are only a few studies of free dust collisions where both collision partners are larger than 1 cm \citep{beitz2011, schraepler2012}. Other collision modes at larger sizes studied, e.g. include one solid impactor, colliding with a dusty medium \citep{colwell2008}. In all these collisions, there is a lot of fragmentation, but mass gain in certain low velocity regimes below 2 m s$^{-1}$ is also possible.\\
Previous experiments on the collision dynamics of targets larger than 10 cm using planetesimal analogs were carried out using chemically bond material such as gypsum \citep{setoh2007, okamoto2009}. The dust agglomerates used here are solely bound by surface forces. This means that no collision experiments exist with two colliding dust aggregates being larger than 10 cm. The problem is that an increase in size by one order of magnitude corresponds to an increase in mass of three orders and decimeter aggregates consist of kilogram-masses. This is at the limit of handling and to avoid a bias induced by gravity such experiments need to be carried out at microgravity. This paper is the first study to close this gap of larger particle collisions. It is likely approaching a size limit accessible in the laboratory. Therefore there is some effort to describe collisions of larger aggregates by numerical simulations \citep{geretshauser2011a, geretshauser2011b}.\\
Decimeter to meter bodies are special as they couple to the gas on an orbital timescale and drift inward the fastest.\\
According to \citet{weidenschilling1977} the radial drift velocity depends on the aggregate size and reaches a maximum of 1 AU in 100 years for bodies of decimeter to meter size. Larger bodies are increasingly less affected by gas drag as their mass becomes larger and their motion more inert. To reach planetesimal size (kilometer), this critical size range has to be overcome. Either coagulation has to be faster than the drift timescales, or additional processes lead to a fast formation of larger bodies.\\
This leads to a second group of growth models that are based on gravitational attraction of solid particles with sufficiently large particle density. Some models are based on particle concentration, e.g. by baroclinic instabilities \citep{lyra2011}. Here, particles are trapped within stable vortices. More turbulent eddies with turbulence supposed to be generated by magnetorotational instabilities also tend to concentrate decimeter particles or particles of Stokes numbers close to 1 \citep{johansen2006, lambrechts2012}. Within only a few orbits, particle densities might overcome the density threshold for gravitational collapse. Another mechanism to clump particles is the streaming instability \citep{Youdin2005}.\\
Independent of the detailed source for the concentration of solids, decimeter bodies play a crucial role for these models due to the fact that they are on the verge to decouple from the surrounding gas. The concentration mechanisms described in these models are therefore strongest for decimeter sized aggregates (depending on the location within the disk). A review of these models is given by \citet{chiang2010}. Collisions of decimeter particles are also important in the concentration and instability models. A fragmentation might decrease the size so that the fragments are no longer following the concentration and might escape if not reaccreted by close by particles.\\
The collision velocity / kinetic energy threshold for fragmentation is one important parameter in the context of planetesimal formation. First studies on the threshold conditions for fragmentation and bouncing of macroscopic dust agglomerates were presented by \citet{beitz2011}. They performed experiments with dust spheres and dust cylinders with a typical diameter of 3 cm. To characterize the outcome of collisions between dust aggregates they defined the fragmentation strength of a collision as $\mu = M_f/M_0$, with the mass of the original aggregate $M_0$ and the mass of the largest fragment $M_f$. In case of growth $\mu$ gets larger than 1, as the largest resulting part is the grown aggregate. Collisions with $\mu$ = 0.5 are characterized as catastrophic disruption. They showed that the critical specific kinetic energy (kinetic energy, normalized by the mass) at the threshold between bouncing and fragmentation, i.e. $\mu $=1, for these samples is at $Q_{\mu =1} = 10^{-2}$ J kg$^{-1}$, which corresponds to a typical collision velocity of about 0.2 m s$^{-1}$, and predicted the following scaling law:
\begin{equation}
Q_{\mu =1} = \left(\frac{r}{3.4\cdot 10^{-5}\,\mathrm{m}}\right)^{-0.95} \mathrm{J} \mathrm{kg}^{-1}\;.
\label{scale}
\end{equation}
Here, $r$ is the particle radius (in meters) and $Q_{\mu =1}$ is the critical specific kinetic energy of the aggregates. More kinetic energy leads to fragmentation, less to bouncing without any mass transfer. In a later study \citet{schraepler2012} extended the studied size range to cylinders of 5 cm diameter, which corresponds to a typical aggregate mass of about 120 g. For larger aggregates only theoretical studies exist so far. For macroscopic dust agglomerates only smoothed particle hydrodynamics (SPH)-simulations have been performed \citep{geretshauser2011a, geretshauser2011b}. Although the SPH-code used has been calibrated for millimeter particles, experimental verification for large dust agglomerates is necessary. Here, we present the first experimental data on collisions between dust aggregates in the decimeter range with a mass one order of magnitude larger than in all previous studies.\\

\section{Experiment}

\subsection{Sample Preparation}

Previous studies showed that the mechanical properties of dust agglomerates are determined by the size distribution of the dust particles and aggregate porosity
\citep{meisner2012, blum2006, schraepler2012, blum2008}. In this study quartz powder is used, which consists of irregular grains in the size range of 0.1 to 10 $\mathrm{\mu m}$, with 80\% of the mass in the range of 1 - 5 $\mathrm{\mu m}$ (producer: Sigma-Aldrich).\\
In order to build large aggregates the dust is pressed into a cylindrical mould and then pushed out by the same mechanism. In this way we obtain cylindrical agglomerates with diameter of 12 cm, a height of about 12 cm and a mean volume filling factor of \mbox{$\phi$ = 0.437$\pm$}0.004. Each aggregate has a mass of about 1.5 kg. All aggregates were weighed prior to the collision experiment and their height was measured to determine their exact volume and their volume filling factor.\\
Due to technical reasons (maximum pressure) it is not possible to produce spherical or irregularly shaped dust aggregates of this size and volume filling. \citet{beitz2011} investigated the collision dynamics of dust agglomerates with a similar volume filling factor with different shapes (spheres and cylinders). They show that the accretion efficiency depends on the volume filling factor and not on the shape of the agglomerate. Their results indicate that this is also the case for the fragmentation threshold and coefficient of restitution. \citet{meisner2012} show that the mechanical properties of dust agglomerates, that is compressive and tensile strength, strongly depend on the volume filling factor. \citet{setoh2007} also found no dependency of the impact strength $Q^*$ on the projectile shape. They investigated the collisions of porous sintered glass beads. Moreover particles in a protoplanetary disk are usually irregularly shaped. Such particles will mostly collide at one point of their surface and are thereby close to the collision geometry of spheres.\\
In addition to that, \citet{beitz2011} found no dependency of the coefficient of restitution on the impact parameter.\\
Also variations in volume filling were not feasible in this size range. For more compact bodies the maximum pressure of the press was not sufficient, whereas more porous aggregates are too weak to be handled properly at the drop tower. The obtained mean volume filling of $\phi =$ 0.437$\pm$0.004 is larger than the volume filling measured for growing dust aggregates \citep{meisner2012, teiser2011}, but corresponds well to previous collision studies by \citet{beitz2011} and \citet{schraepler2012}. The modulus of elasticity and the tensile strength of compact dust agglomerates made of the same Quartz powder have been determined by \citet{meisner2012}. For a volume filling of $\phi$ = 0.44 they found a tensile strength of \mbox{$\sigma = 2.2 \cdot 10^3$} Pa, a modulus of elasticity of \mbox{$E = 1.79 \cdot 10^7$} Pa and a compression stress of 55 kPa.

\subsection{Setup for the Collision Experiments}

\begin{figure}[hb]
\epsscale{1.}
\plotone{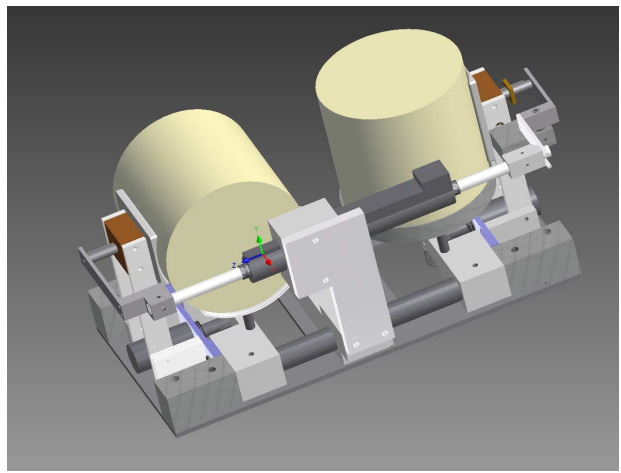}
\caption{Experimental setup for the collisions of the decimeter dust aggregates \label{aufbau}}
\end{figure}

The collision experiments are carried out in the drop tower in Bremen, where microgravity conditions ($< 10^{-6}g$) are provided for 4.7 s. The experimental setup (see Fig.~\ref{aufbau}) is placed inside a vacuum chamber at a pressure of $p \lesssim \mathrm{10^{-2} mbar}$. The residual air within the pores of the dust agglomerates as well as the influence of gas drag on moving aggregates is therefore negligible.\\
The dust aggregates are placed on two sample mounts. One of the aggregates is standing upright, and the other one is lying on one side. This is as close as possible to the geometry of a collision between two spherical bodies. Even for irregularly shaped agglomerates the most probable impact configuration features only a small contact area.\\
\\
When the drop capsule reaches microgravity conditions the sample mounts are pulled down by two solenoids. The two dust aggregates are then accelerated by a linear motor via two pistons. The pistons stop, once they have reached the pre-adjusted velocity and the collision takes place, when both dust aggregates are completely free. The collision of the aggregates is observed by two high speed cameras at 500 frames per second and illuminated by two LED systems. The two cameras are positioned perpendicular to one another to obtain full three dimensional information. The two LED systems are mounted close to the two cameras, so that the critical part of the dust aggregates is illuminated well with no disturbing shadows. One camera observes the collision from the top, the other one from the front.

\section{Results}
\begin{figure}[ht]
\epsscale{1.}
\plotone{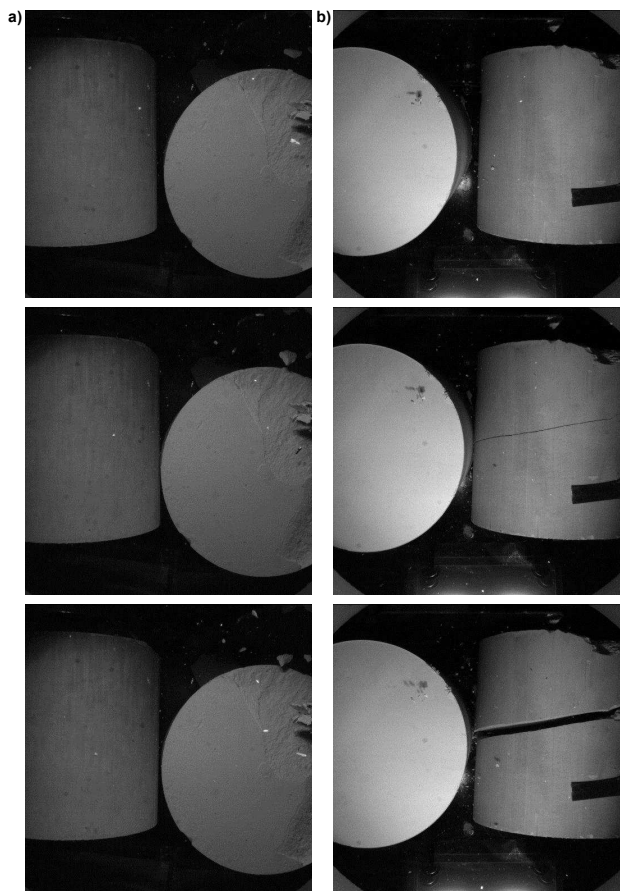}
\caption{Examples for the two different collision outcomes: (a) bouncing, (b) fragmentation\label{koll}}
\end{figure}

Fig. \ref{koll} shows examples for bouncing and fragmentation. Each camera provides information on linear motion and rotation of the two aggregates.\\

\subsection{Critical Fragmentation Velocity}
The linear motion is measured by tracking the face of the aggregate, which is seen face-on, and by tracking the mantle (edge) of the other aggregate. This is done with the images of both cameras. Features on the dust surfaces (mantle as well as face) are used to gain additional data on the rotation of the bodies.\\
Altogether the linear motion and two perpendicular components of the rotation are derived. This leads to a kinetic energy of a dust aggregate of  $E_{\mathrm{kin}} = \frac{1}{2} \left( m v^2 + I_x \omega_x^2+I_y \omega_y^2\right)$ where $I_x$ is the moment of inertia around the symmetry axis, $I_y$ is the moment of inertia around the transverse axis and $m$ is the mass. This is done for all aggregates before and after a collision. In the case of fragmentation, the volume of the fragments is derived from the camera images.\\
In all experiments, in which fragmentation could be observed, the aggregates broke exactly into two cylindrical pieces with a smooth fracture layer. The volume of these fragments could therefore easily be derived from the camera images. In two experiments a small part of one aggregate chipped off prior to the experiment while handling the drop capsule. For these experiments the volume of the remaining aggregate is also derived from the camera images.\\

\begin{figure}[hb]
\epsscale{1.}
\plotone{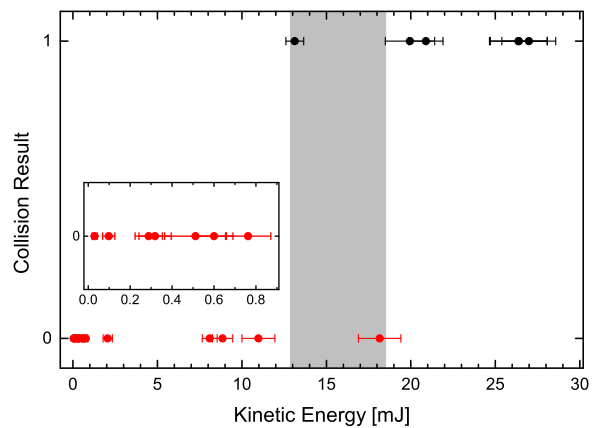}
\caption{Results of the collisions (bouncing = 0, fragmentation = 1 \label{grenz}}
\end{figure}

The error of the linear and angular velocity of an agglomerate was calculated with the method of error propagation, assuming an error of one pixel for the aggregate position. Combined with the errors in the calculation of mass ($\Delta m$=0.5g) and volume ($\Delta h$=0.1cm) we can determine the error of the kinetic energy of an agglomerate.
Fig.~\ref{grenz} shows the results of the collisions, with bouncing set to 0 and fragmentation set to 1, plotted versus the sum of the kinetic energies of the two aggregates prior to the collision. The small inset shows a more detailed view of collisions at low kinetic energies. Fragmenting collisions where only one of the aggregates breaks up were observed as well as collisions where both aggregates break up.\\
The pistons, which accelerate the dust aggregates stop in a fixed position once they reach their pre-adjusted velocity. In the case of a bouncing collision, the aggregates can collide with the now fixed pistons. As these collisions do not lead to fragmentation, the aggregates bounce off and can collide with each other for a second time. In this case one experiment can lead to a maximum of four bouncing collisions with decreasing impact velocities and very small impact velocities are reached.\\
The  boundary between bouncing and fragmentation lies in the area highlighted in gray. It is determined by the largest kinetic energy for which bouncing was observed and the smallest energy at which fragmentation occurred. The center of the marked transition zone corresponds to an impact velocity of \mbox{$v$ = 16.2 cms$^{-1}$.} This can be characterized as the threshold velocity for fragmentation.

\subsection{Coefficient of Restitution}

\begin{figure}[ht]
\begin{center}
\epsscale{1.}
\plotone{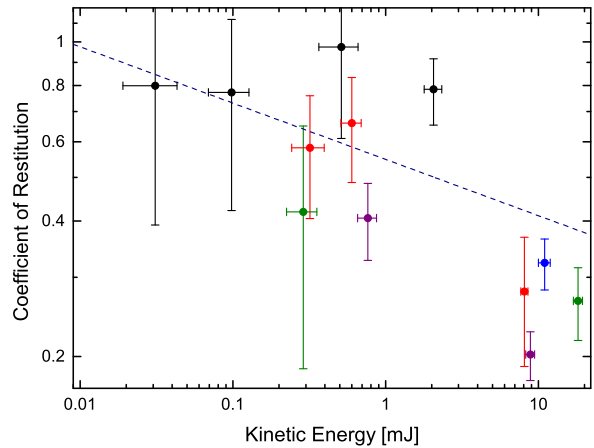}
\caption{Coefficient of restitution: symbols in the same color belong to multiple collisions within a drop tower experiment.\label{rest}}
\end{center}
\end{figure}

We here define the coefficient of restitution $e$ as the ratio of the total kinetic energy after and before the collision. In this way, the rotation of the aggregates before and after the collision is included in the calculation, as done by \citet{schraepler2012}. To calculate the coefficient of restitution only the bouncing collisions are taken into account. The results are shown in Fig~\ref{rest}. The error bars shown in the figure were calculated using the method of error propagation. As already mentioned before, one aggregate can perform multiple collisions within one experiment due to collisions with the fixed pistons. Therefore in Fig.~\ref{rest} the symbols in different colors stand for different aggregates. A clear trend is visible that the coefficient of restitution is significantly lower for larger collision velocities.\\
\section{Discussion}

\subsection{Threshold Conditions}


In the study by \citet{beitz2011} the fragmenting dust aggregates usually break up into several fragments with different sizes. This is different for the collisions observed in this work. Here, the fragmenting aggregate breaks up into exactly two pieces. The fracture layer starts exactly at the contact point between both colliding bodies. The fragment size therefore depends on the impact parameter.\\
For a perfectly central collision, both fragments have the same size (see Fig. \ref{koll}). In the case of non central collisions, the size ratio depends on the exact position of the contact point. The fragmentation characteristic is the same for all experiments, which leads to the conclusion that only fragmentation directly at the fragmentation limit has been observed. The fragmentation strength therefore can be treated as $\mu$ = 1 for all collisions in which fragmentation occurred.\\
For collisions between dust aggregates \citet{beitz2011} found a dependency between the critical specific kinetic energy and the aggregate size and proposed the scaling law of equation \ref{scale}. For the fragmentation threshold ($\mu$ = 1) for decimeter bodies this corresponds to a specific energy of $Q_{\mu =1} \approx 8 \cdot 10^{-4} \mathrm{J \, kg^{-1}}$. The experiments in this study give a critical kinetic energy of $E_{\mathrm{kin}} = 15.7\, \mathrm{mJ}$, which corresponds to an impact velocity of $16.2\, \mathrm{cms^{-1}}$ and a specific kinetic energy of $Q_{\mu =1} = 5 \cdot 10^{-3} \mathrm{J\,kg^{-1}}$. This is almost one magnitude larger than the values predicted by \citet{beitz2011}.\\
Due to technical reasons the velocity range within this study is limited, so the fragmentation strength for $\mu$ = 0.5 could not be determined. For smaller aggregates \citet{beitz2011} show that $\mu$=1 and $\mu$=0.5 have the same scaling law for the dependency of the critical specific energy on the agglomerate size. Due to the limitations in the velocity range we cannot confirm the validity of this scaling law for decimeter agglomerates. Assuming it is valid here, the critical specific energy for $\mu$~=~0.5 would be about $Q_{\mu =0.5} = 5 \cdot 10^{-2} \mathrm{J\, kg^{-1}}$.\\

\subsection{Coefficient of Restitution}

The coefficient of restitution clearly depends on the kinetic energy of the colliding dust aggregates, which corresponds to the results of smaller aggregates by \citet{schraepler2012}. The coefficient of restitution is about 0.27 for energies close to the fragmentation threshold. For very small kinetic energies collisions become more elastic. Obviously a minimum energy is needed to trigger inelastic restructuring within the dust aggregates. The restructuring (compression) is below the resolution of the camera images, so it cannot be characterized in more detail.\\
For smaller dust aggregates \citet{schraepler2012} found a power law, $e \propto v^{-1/4}$ for low velocities, describing the dependency between the coefficient of restitution and the collision velocity, according to a solid state model by \citet{thornton1998}.\\
The power law $e \propto v^{-1/4}$ corresponds to $e \propto E^{-1/8}$ and is shown in Fig.~\ref{rest} (blue dashed line) in order to compare the experimental data to a theoretical model. The experimental data fit quite well to the curve, but a real fit of the data is not possible with this limited dataset. The general behaviour is in good agreement with the findings of \citet{schraepler2012}. In comparison to the experiments of \citet{schraepler2012} the dust aggregates used here have a mass that is a factor of 10 higher.\\
In the study by \citet{beitz2011} no significant dependency of the coefficient of restitution from the collision velocity has been detected for more compact dust aggregates ($\phi \approx$ 0.5). One possible reason for this is their use of more compact dust aggregates.\\

\subsection{Astrophysical Applications}
The results of the collision experiments are interesting for several different astrophysical applications, in which collisions between decimeter sized aggregates occur.
\begin{itemize}
\item \emph{Disks, collisional evolution:} According to various disk models, the expected typical collision velocities for decimeter bodies exceed the fragmentation threshold by far. These velocities are of the order of 10 m s$^{-1}$ or even larger, for protoplanetary disks \citep{brauer2008, weidenschilling1993, desch2007, chiang2010} as well as for late debris disks \citep{krivov2000}. In both scenarios decimeter bodies will be destroyed by mutual collisions. Experiments show that further growth might still be possible, as small particles can effectively be collected by large bodies \citep{teiser2011}.\\
Observations of protoplanetary disks show the presence of small dust grains \citep{Dullemond2005, perez2012}. The theoretical calculations of \citet{Zsom2010} predict a bouncing barrier for mutual collisions of millimeter sized bodies. For decimeter bodies there is no bouncing barrier as mutual collisions at the presumed velocities in protoplanetary disks lead to fragmentation. Calculations of \citet{windmark2012a, windmark2012b} consider a Maxwellian velocity distribution for collisions. In this scenario some particles are able to overcome the critical size range by sweeping up small dust grains. This emphasizes the importance of small fragments that are presumably produced by collisions of decimeter bodies.
\item \emph{Disks, gravitational instabilities:} All current models based on gravitational instabilities propose hydrodynamic processes, which form local regions (vortices) in which solid particles can be trapped. Especially baroclinic instabilities \citep{lyra2011} and streaming instabilities \citep{Youdin2005} can concentrate solid particles in clusters and relative velocities are kept low, as all particles follow the movement of the cluster. In this scenario relative velocities might stay small, so at least a significant part of the large (decimeter) dust aggregates might survive.
\item \emph{Planetary rings:}
Studies on the rings of Saturn show that the rings mainly consist of water ice \citep{nicholson2008}, but also contain regolith covered ice particles \citep{elliott2011, filacchione2012}. The physical properties of ice particles, especially the fragmentation conditions, differ significantly from those of dust agglomerates. The coefficient of restitution of a collision however is mainly determined by the surface properties \citep{bridges2001}. This means that the regolith layer can be the dominant factor here. In this way the coefficient of restitution of dust agglomerates and regolith covered ice particles is similar. The relative velocities of the ring particles are of the order of a few cm s$^{-1}$ \citep{porco2008} and are thereby in the range investigated in our experiments.\\
Several studies describe planetary rings as equilibrium systems, in which encounters between ring bodies trigger relative velocities (with a chaotic velocity distribution), while non-elastic collisions dissipate kinetic energy. This requires a decreasing coefficient of restitution for increasing collision velocities, which leads to a direct dependency between the coefficient of restitution and the optical depth \citep{goldreich1978}. For Saturn's main rings a coefficient of restitution of $e \gtrsim$ 0.6 has been proposed by \citet{goldreich1978}. Our experiments show that this is feasible for decimeter dust aggregates, if the collision velocities are small enough.\\
\end{itemize}
\section{Conclusions}
The critical fragmentation velocity for collisions of cylindrical decimeter sized aggregates lies at about $16.2\,\mathrm{cm s^{-1}}$ (aggregates are made up of irregular quartz grains and have a volume filling factor of $\phi$ = 0.437$\pm$0.004). The aggregates behave elastically at low velocities, while the coefficient of restitution decreases with increasing collision velocity. Bouncing collisions close to the fragmentation threshold reduce the ability for elastic collisions.\\
These results can be applied to various astrophysical environments. The fragmentation threshold is much lower than the expected velocities for mutual collisions of decimeter aggregates in protoplanetary disks. Even then growth of aggregates might still be possible as small particles can be swept up by the few bigger aggregates that overcome the critical size range \citep{windmark2012a, windmark2012b}. The results are also interesting for the study of planetary rings as the collision velocities are supposed to be in the same range investigated here. The decreasing coefficient of restitution fits quite well to the description of the planetary rings as equilibrium systems \citep{goldreich1978}.


\acknowledgments

We would like to thank Caroline De Beule and Janine van Eymeren for their invaluable help in carrying out the drop tower experiments.\\
Furthermore we would like to thank the Deutsche Forschungsgemeinschaft (DFG) for their funding within the frame of the SPP 1385. Access to the drop tower was granted by the European Space Agency as "CODE~I".

\bibliography{literature}

\end{document}